\documentclass[fleqn,10pt]{wlscirep}
\usepackage[utf8]{inputenc}
\usepackage[T1]{fontenc}
\usepackage{lineno}
\usepackage{float}

\title{A diffusion MRI tractography atlas for concurrent white matter mapping across Eastern and Western populations}

\author[1,$\dag$]{Yijie Li}
\author[1,$\dag$]{Wei Zhang}
\author[2]{Ye Wu}
\author[3]{Li Yin}
\author[1]{Ce Zhu}
\author[4]{Yuqian Chen}
\author[5]{Suheyla Cetin-Karayumak}
\author[5]{Kang Ik K Cho}
\author[4]{Leo R. Zekelman}
\author[5,6]{Jarrett Rushmore}
\author[5]{Yogesh Rathi}
\author[5]{Nikos Makris}
\author[4,*]{Lauren J. O’Donnell}
\author[1,*]{Fan Zhang}
\affil[1]{School of Information and Communication Engineering, University of Electronic Science and Technology of China, Chengdu, China}
\affil[2]{School of Computer Science and Engineering, Nanjing University of Science and Technology, Nanjing, China}
\affil[3]{West China Hospital of Medical Science, Sichuan University, Chengdu, China}
\affil[4]{Department of Radiology, Brigham and Women’s Hospital, Harvard Medical School, Boston, USA}
\affil[5]{Department of Psychiatry, Brigham and Women’s Hospital, Harvard Medical School, Boston, USA}
\affil[6]{Department of Anatomy and Neurobiology, Boston University School of Medicine, Boston, USA}

\affil[*]{Lauren J O’Donnell (odonnell@bwh.harvard.edu) and Fan Zhang (fan.zhang@uestc.edu.cn) are co-corresponding-authors.}

\affil[$\dag$]{Yijie Li and Wei Zhang are co-first-authors.}

\begin{abstract}

The study of brain differences across Eastern and Western populations provides vital insights for understanding potential cultural and genetic influences on cognition and mental health. Diffusion MRI (dMRI) tractography is an important tool in assessing white matter (WM) connectivity and brain tissue microstructure across different populations. However, a comprehensive investigation into WM fiber tracts between Eastern and Western populations is challenged due to the lack of a cross-population WM atlas and the large site-specific variability of dMRI data. This study presents a dMRI tractography atlas, namely the \textit{East-West WM Atlas}, for concurrent WM mapping between Eastern and Western populations and creates a large, harmonized dMRI dataset (n=306) based on the Human Connectome Project and the Chinese Human Connectome Project. The curated WM atlas, as well as subject-specific data including the harmonized dMRI data, the whole brain tractography data, and parcellated WM fiber tracts and their diffusion measures, are publicly released. This resource is a valuable addition to facilitating the exploration of brain commonalities and differences across diverse cultural backgrounds.

\end{abstract}
\begin{document}

\flushbottom
\maketitle

\thispagestyle{empty}

\section*{Background \& Summary}

Differences in genetics, cultural backgrounds, and environmental influences play a vital role in shaping the structure and function of the human brain\cite{tooley2021environmental,ge2023increasing}. There is a surge of interest in studying the brains of Eastern and Western populations to investigate potential cultural influences on cognition and mental health\cite{ge2023increasing,kochunov2003localized,nisbett2005influence,chua2005cultural,han2014cultural,ge2015cross,gao2022cultural}. Noninvasive neuroimaging is an important tool to uncover the neural basis of the human mind and behavior. In particular, magnetic resonance imaging (MRI) has greatly improved the understanding of cross-cultural brain differences in morphology, cortical thickness, and other aspects between Eastern and Western populations\cite{han2014cultural,ge2015cross,gao2022cultural,tang2010construction,yang2020sample,kang2020differences,wei2023native,tang2018brain,huang2019culture}.

Diffusion MRI (dMRI) is an advanced MRI technique that can probe the diffusion of water molecules in biological tissues to characterize the underlying microstructure\cite{basser1994mr}. In particular, dMRI enables a computational process, namely tractography, that uniquely enables in-vivo reconstruction of the brain’s white matter (WM) connections at macro scale\cite{basser2000vivo}. Quantitative approaches using tractography have become popular tools for studying the brain's connectivity and tissue microstructure, including connectome-based analyses to investigate the structural connectivity of the entire brain and tract-specific analyses to investigate particular anatomical fiber tracts\cite{zhang2022quantitative}. Currently, dMRI tractography plays a prominent role in studying the brain’s WM connections in health and disease\cite{zhang2022quantitative,piper2014application,pannek2014magnetic,essayed2017white}. 

Multiple studies have utilized dMRI tractography to explore cross-cultural brain studies between Eastern and Western populations. Existing studies focus on connectome-based analyses\cite{ge2023increasing,zhang2019structural,suo2021anatomical}, which have revealed distinct brain topographic features specific to each population in terms of whole brain structural connectivity strengths. However, investigation of specific anatomical fiber tracts (e.g., the arcuate fasciculus and the corticospinal tract) between Eastern and Western populations is lacking. Unlike connectome-based analyses that are data-driven to explore the structural connectivity of the entire brain, tract-specific analyses enable hypothesis-driven studies of certain anatomical fiber tracts\cite{zhang2022quantitative}. This allows detailed investigation of local WM regions in association with certain brain functions, an approach that has been widely used in neuroscientific and clinical research studies\cite{yeo2014different,alexander2007diffusion,shany2017diffusion,zekelman2022white,ribeiro2024white}. Yet, a comprehensive investigation into WM fiber tracts between Eastern and Western populations is still missing due to the following challenges.

The first challenge is the lack of dMRI tractography atlases that enable concurrent mapping of WM fiber tracts between Eastern and Western populations. In neuroimaging research, brain atlases serve as essential tools for standardized representation of the anatomical structures and functional regions within the brain\cite{laird2009ale,mazziotta2001probabilistic,collins1994automatic}. Creating a brain atlas generally involves constructing a common template derived from a group of individuals, which can be applied to diverse individuals to identify subject-specific anatomically segregated brain regions. Previous studies examining the brains across different cultures, focusing on the cortical surface and brain morphology\cite{yang2020sample,yang2020constructing,liang2015construction}, have highlighted the necessity of developing atlases specific to each population. This is crucial to accurately capture subtle anatomical variations that exist within a particular cultural group. In the dMRI tractography literature, existing WM atlases primarily originate from data collected from Western populations\cite{catani2008diffusion,yendiki2011automated,roman2017clustering,zhang2018anatomically,yeh2018population,radwan2022atlas}. While these atlases can be directly applied to data from an Eastern population, they might not capture the subtle inter-population anatomical variability of the WM\@. Therefore, there is a high demand for an across-population WM atlas that facilitates concurrent mapping of WM fiber tracts between Eastern and Western populations, promoting more inclusive cross-culture brain studies.

A second challenge in studying cross-cultural brain WM differences between Eastern and Western populations is the site-specific variability in dMRI data. Unlike other types of group-wise analyses (e.g., health vs disease) where neuroimaging data of the participants can be acquired on the same MRI scanner from a single acquisition site, cross-cultural analyses usually rely on data acquired at multiple sites from different countries or continents. In dMRI, it is well known that variations in site-specific data arising from different MRI scanners and/or acquisition protocols can result in significant biases in measures of WM connectivity and microstructure\cite{vollmar2010identical,mirzaalian2016inter,karayumak2019retrospective,cetin2018harmonizing,liu2020ms,ning2020cross,tax2019cross}. Thus, many studies underscore the significance of eliminating site-specific biases before running joint analysis in multi-site studies\cite{mirzaalian2016inter,karayumak2019retrospective,mirzaalian2018multi,pinto2020harmonization,mirzaalian2015harmonizing,fortin2017harmonization,huynh2019multi,moyer2020scanner,hagler2019image}.\ dMRI harmonization is an effective way to mitigate measurement differences attributed to scanner-, protocol-, or other site-related effects, and it has been used in recent studies to enable pooled dMRI data analyses\cite{mirzaalian2016inter,karayumak2019retrospective,mirzaalian2018multi,pinto2020harmonization,mirzaalian2015harmonizing,fortin2017harmonization,huynh2019multi,moyer2020scanner}. However, there are currently no harmonized dMRI resources to enable joint dMRI analysis between Eastern and Western populations. 

In light of the above, this study presents a novel dMRI tractography atlas for concurrent WM mapping across Eastern and Western populations, referred to as the \textit{East-West WM Atlas}. High-quality dMRI data from the Human Connectome Project (HCP)\cite{van2013wu} and the Chinese Human Connectome Project (CHCP)\cite{ge2023increasing}, acquired from Western and Eastern populations, respectively, are used for the atlas creation. To remove site-specific biases from the multi-site data, we employ our advanced dMRI harmonization algorithm\cite{karayumak2019retrospective,mirzaalian2018multi} that reconciles the raw dMRI signals across the two datasets while preserving inter-population variability. This generates a pooled large-scale dMRI dataset including a total of 306 individuals (153 for each cultural population). From this dataset, we create a cross-population WM atlas using our well-established fiber clustering pipeline\cite{zhang2018anatomically,o2012unbiased,o2007automatic} that enables simultaneous tractography atlas creation across populations. This generates an anatomically curated atlas including a tract-level parcellation of major anatomical fiber tracts as well as a fine-scale parcellation of the entire WM. Importantly, the created atlas, together with the harmonized datasets, enables subject-specific WM parcellation for direct comparison of the WM connectivity and microstructural measures between the two populations. Constituting an essential new resource for cross-cultural brain studies, all code of related computational tools and the data including the harmonized dMRI dataset, the created WM atlas, the subject-specific WM tracts, and the associated diffusion measures are openly available. 

\section*{Methods}

\begin{figure}[H]
\centering
\includegraphics[width=\linewidth]{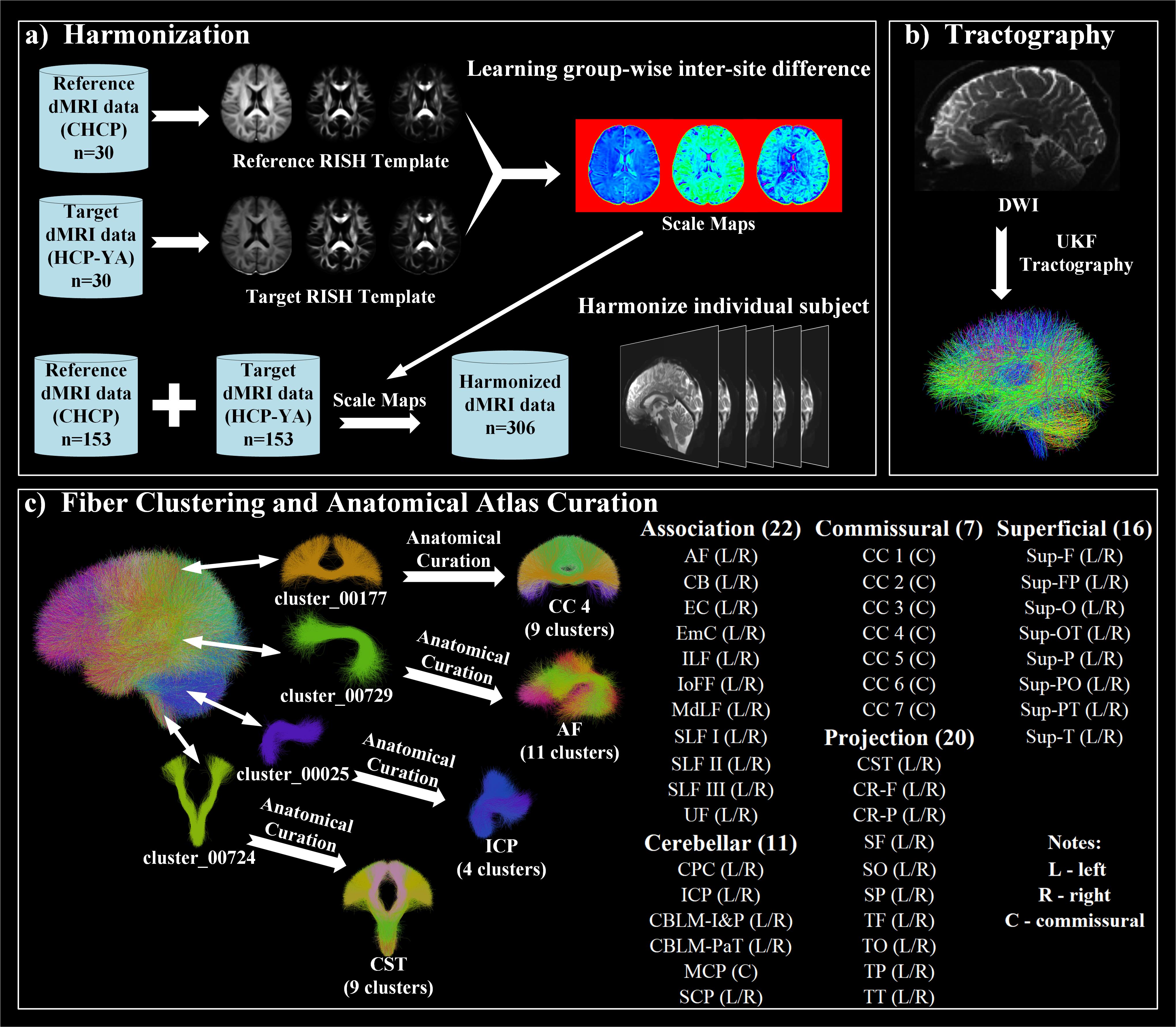}
\caption{Method overview: (a) dMRI data harmonization to eliminate inter-site discrepancies between the CHCP and HCP datasets. (b) Whole brain tractography for the reconstruction of WM connections from the entire brain. (c) Group-wise fiber clustering atlas creation concurrently using tractography data from the CHCP and HCP populations. A total of 76 anatomical fiber tracts categorized into 5 tract categories are included. }
\label{fig:MethodOverview}
\end{figure}

This section introduces the dMRI dataset and the computational steps for developing the proposed WM atlas (See Figure \ref{fig:MethodOverview} for a method overview).

\subsection*{Datasets and Participants}

This study utilized dMRI datasets from two sources: the Human Connectome Project Young Adult (HCP-YA) database, obtained from participants living in a Western culture from the USA\cite{van2013wu}, and the Chinese Connectome Project (CHCP) database, obtained from participants living in an Eastern culture from China\cite{ge2023increasing}. The HCP-YA database comprised dMRI data from over 1,000 young healthy adults (22 to 37 years old), while the CHCP database included dMRI data from over 300 healthy adults including teenagers, young adults, and elderly adults (18 to 79 years old). In our investigation, we used two subsets of participants from CHCP and HCP-YA. We included the young adult participants (22 to 37 years) from the CHCP dataset and selected a subset of HCP-YA participants with similar age and sex distributions. Overall, in our study, each dataset contained data from 153 participants (HCP-YA: $24.2\pm1.4$ years, 72 females and 81 males; CHCP: $23.9\pm2.4$ years, 68 females and 85 males; no significant age or sex differences), resulting in a large cohort of 306 participants used for the atlas creation and technical validation. 

\subsection*{MRI Acquisitions and Preprocessing}

The dMRI data of the above participants were publicly available in the HCP-YA (\url{db.humanconnectome.org}) and CHCP (\url{doi.org/10.11922/sciencedb.01374}) databases. The detailed acquisition parameters are described in \cite{van2013wu} and \cite{ge2023increasing}. Briefly, the dMRI data in HCP-YA was acquired on a customized 3T Connectome Siemens Skyra scanner, with the following acquisition parameters: $TE\ =\ 89.5\ ms$, $TR\ =\ 5520\ ms$, voxel size = 1.25 $\times$ 1.25 $\times$ 1.25 mm$^3$, and a total of 288 volumes including 18 baseline images and 90 diffusion-weighted images at each of the three shells of $b\ =\ 1000/2000/3000\ s/mm^3$. The dMRI data in CHCP was acquired on a 3T Siemens Prisma MRI scanner with the following acquisition parameters: $TE\ =\ 86\ ms$, $TR\ =\ 3500\ ms$, voxel size = 1.5 $\times$ 1.5 $\times$ 1.5 mm$^3$, 14 baseline images at b = 0, 93 diffusion-weighted images at $b\ =\ 1000\ s/mm^3$, and 92 diffusion-weighted images at $b\ =\ 2000\ s/mm^3$. In both dMRI datasets, we used only single-shell dMRI data for reasonable computation time and memory use when performing tractography. We employed the $b\ =\ 2000\ s/mm^3$ data as these were available in both databases, and the angular resolution is better and more accurate at higher b-values\cite{descoteaux2007regularized,ning2015sparse}.

Both the HCP-YA and CHCP datasets were preprocessed with standard dMRI processing steps, including brain masking, eddy current-induced distortion correction, motion correction, and EPI distortion correction. Specifically, for the HCP-YA data, we used the dMRI data already preprocessed with the HCP minimum processing pipeline\cite{glasser2013minimal}. For the CHCP data, we used our well-established dMRI data processing pipeline\cite{zhang2018anatomically} (\url{https://github.com/pnlbwh/pnlpipe/tree/v2.2.0}), including: CNN brain masking\cite{cetin2024harmonized} (\url{https://github.com/pnlbwh/CNN-Diffusion-MRIBrain-Segmentation/tree/v0.3}) for dMRI images, and eddy current, motion, and EPI distortion correction using  FSL topup and eddy tools  (version 6.0.6.5)\cite{jenkinson2012fsl}.

\subsection*{Data Harmonization}

While the HCP-YA and CHCP dMRI datasets underwent similar preprocessing, there were inherent scanner-specific biases due to the involvement of multiple acquisition sites, scanners, and acquisition protocols. To mitigate the inter-site variability within this multi-site dMRI study, we employed our retrospective harmonization algorithm\cite{karayumak2019retrospective,mirzaalian2018multi}. Our harmonization method relied on rotation-invariant spherical harmonics (RISH) features at the dMRI signal level. Its primary objective was to remove scanner-specific biases across datasets, accounting for the non-linearities in dMRI data, which may vary by region and tissue. This method was successfully used in multiple studies for dMRI harmonization to enable pooled data analysis\cite{cetin2024harmonized,cetin2023characterization,de2022cross,seitz2022cognitive}.

The harmonization process (\url{https://github.com/pnlbwh/dMRIharmonization/tree/v2.1}) comprised the following two essential steps. Briefly, the first step was to learn dMRI differences between the two datasets while preserving inter-subject biological variability at a group level. To do so, we selected 30 subjects from each of the CHCP and HCP datasets with similar age and sex distributions (no significant differences). For each selected subject, we calculated RISH features to capture orientation-independent microstructural tissue properties and enable the reconstruction of the harmonized dMRI signal\cite{reisert2017disentangling}. Nonlinear mappings of the RISH features were learned from the HCP-YA (target site) to the CHCP data (reference site). While CHCP is selected as a reference site, we have previously shown that the choice of the reference site does not affect the performance of harmonization \cite{karayumak2019retrospective}. The second step of the dMRI harmonization process was to harmonize the dMRI data in each individual subject. This was done by applying the learned mappings to the RISH features computed from the subjects in the target site. This generated harmonized RISH features that correspond to the reference site. Then, from the harmonized RISH features, dMRI signals were reconstructed for the following tractography analyses.

\subsection*{Whole Brain Tractography}

Whole brain tractography was performed using our two-tensor Unscented Kalman Filter (UKF) approach\cite{farquharson2013white,vos2013multi},  as implemented in the ukftractography package (\url{https://github.com/pnlbwh/ukftractography}). The UKF method fitted a mixture model of two tensors to the dMRI data while tracking fibers, employing prior information from the previous step to help stabilize model fitting. UKF was shown to be highly consistent for fiber tracking in dMRI data from independently acquired populations across ages, health conditions, and dMRI acquisitions\cite{zhang2018anatomically,zhang2020deep,xue2023superficial}. 

For each of the harmonized dMRI scans from the HCP-YA and CHCP datasets, we adopted the tractography parameters as used in \cite{zhang2018anatomically} for whole brain WM fiber tracking. In brief, tractography was seeded in all voxels within the brain mask where fractional anisotropy (FA) was greater than 0.1. Tracking stopped where the FA value fell below 0.08 or the normalized mean signal fell below 0.06. To ensure comparability and minimize bias related to variations in streamline counts across subjects and datasets, we uniformly downsampled the whole-brain tractography for each subject to 600,000 streamlines. The diffusion tensors calculated during tractography, together with their FA and mean diffusivity (MD) values, were preserved for subsequent analyses to enable a comprehensive assessment of the WM microstructure. 

\subsection*{Fiber Clustering Atlas Creation}

After obtaining the tractography data per subject, a groupwise whole-brain fiber clustering atlas was created using our robust, data-driven fiber clustering pipeline\cite{zhang2018anatomically,o2012unbiased,o2007automatic}, as implemented in the \textit{whitematteranalysis} (WMA) software (\url{https://github.com/SlicerDMRI/whitematteranalysis}). The pipeline included two key processes: a group-wise tractography registration to align the tractography of all subjects to a common space, and a spectral clustering of tractography to subdivide the registered tractography data into multiple fiber clusters simultaneously. The WMA fiber clustering pipeline has been successfully used for creating WM\cite{zhang2018anatomically} and cranial nerve\cite{zhang2020creation,zeng2023automated} tractography atlases. 

\begin{figure}[H]
\centering
\includegraphics[width=0.75\linewidth]{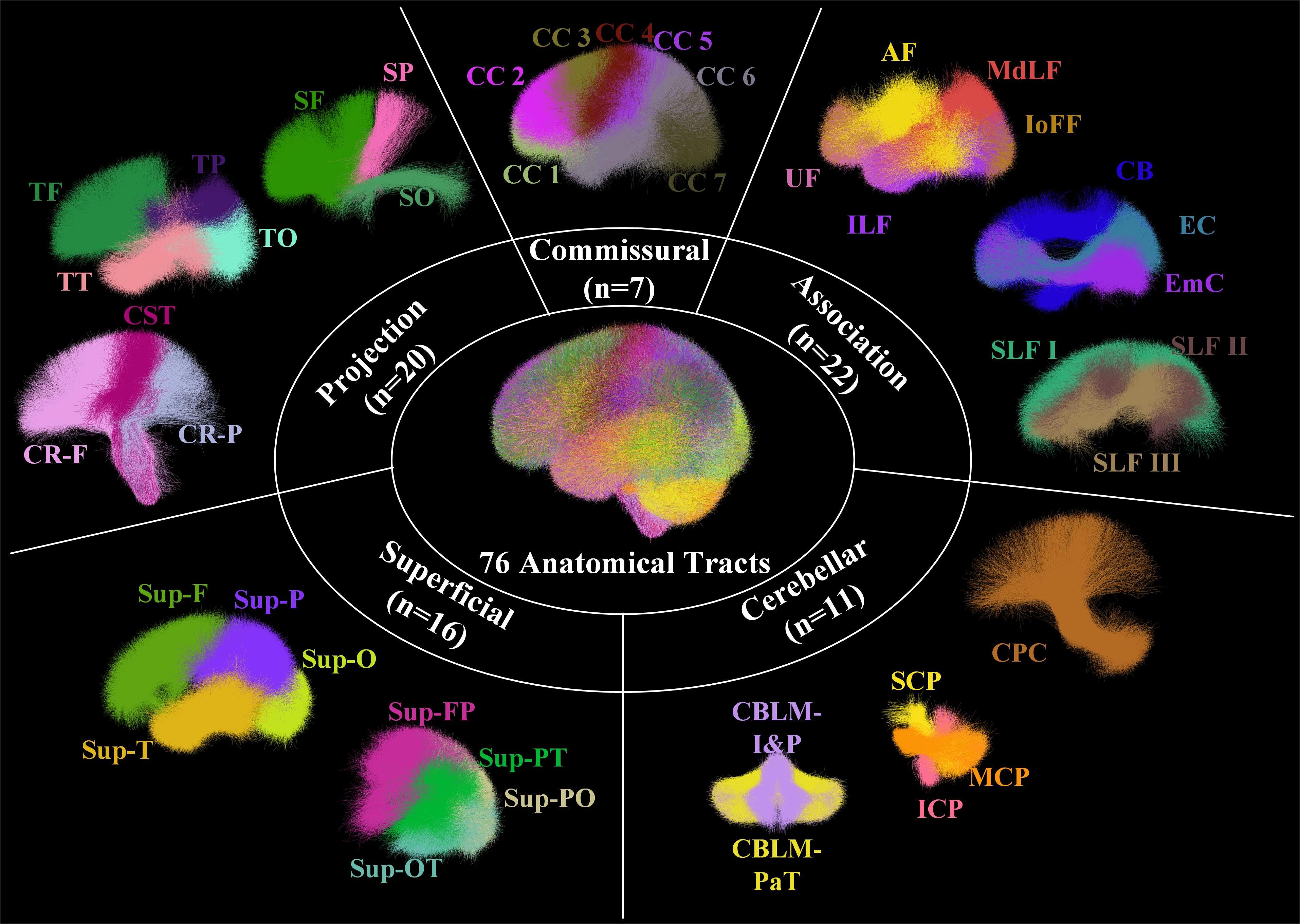}
\caption{Visualization of anatomical fiber tracts included in the proposed atlas, categorized into five tract categories. Association tracts include: arcuate fasciculus (AF), cingulum bundle (CB), external capsule (EC), extreme capsule (EmC), inferior longitudinal fasciculus (ILF), inferior occipito-frontal fasciculus (IoFF), middle longitudinal fasciculus (MdLF), superior longitudinal fascicle I, II, III (SLF I, II, III), and uncinate fasciculus (UF). Cerebellar tracts include: cortical-ponto-cerebellar (CPC), inferior cerebellar peduncle (ICP), intracerebellar input and Purkinje tract (CBLM-I\&P), intracerebellar parallel tract (CBLM-PaT), middle cerebellar peduncle (MCP), and superior cerebellar peduncle (SCP). Projection tracts include: corticospinal tract (CST), corona-radiata-frontal (CR-F), corona-radiata-parietal (CR-P), striato-frontal (SF), striato-occipital (SO), striato-parietal (SP), thalamo-frontal (TF), thalamo-occipital, thalamo-parietal (TP), and thalamo-temporal (TT). Commissural tracts include seven segments of the corpus callosum (CC): rostrum (CC 1), genu (CC 2), rostral body (CC 3), anterior midbody (CC 4), posterior midbody (CC 5), isthmus (CC 6), and splenium (CC 7). Superficial tracts include: superficial-frontal (Sup-F), superficial-frontal-parietal (Sup-FP), superficial-occipital (Sup-O), superficial-occipital-temporal (Sup-OT), superficial-parietal (Sup-P), superficial-parietal-occipital (Sup-PO), superficial-parietal-temporal (Sup-PT), and superficial-temporal (Sup-T).}
\label{fig:TractList}
\end{figure}

In our study, we pooled the tractography data of the 306 subjects from the CHCP and HCP datasets together for atlas creation. This allowed us to learn a tractography atlas for concurrent mapping of the WM connections across the two populations. Specifically, from each subject’s whole brain tractography, 10,000 streamlines were randomly selected, resulting in approximately 3 million streamlines for clustering. The WMA fiber clustering pipeline was used to parcellate all streamlines into K clusters. We chose $K\ =\ 800$, which was shown to be a good parcellation scale of the whole brain WM in previous studies\cite{zhang2018anatomically,o2017automated,wu2021highly}. We then performed anatomical curation of the fiber clusters by annotating each fiber cluster with an anatomical label belonging to a certain anatomical tract (e.g., the corticospinal tract) or an unclassified category. To do so, we leveraged the ORG atlas built only using the HCP data as a reference\cite{zhang2018anatomically,zhang2019test}. We performed co-registration of the two atlases using a tractography-based registration\cite{o2012unbiased} and calculated the mean closest point distances\cite{o2007automatic,moberts2005evaluation} between the clusters in the new atlas and those in the ORG atlas and then assigned each new atlas cluster with the label of the closest ORG cluster. In total, the proposed atlas annotates 60 deep WM tracts and 222 superficial fiber clusters categorized into 16 groups based on their associated brain lobes, as detailed in Figure \ref{fig:TractList}.

\subsection*{WM Parcellation of Individual Subjects}

\begin{figure}[H]
\centering
\includegraphics[width=\linewidth]{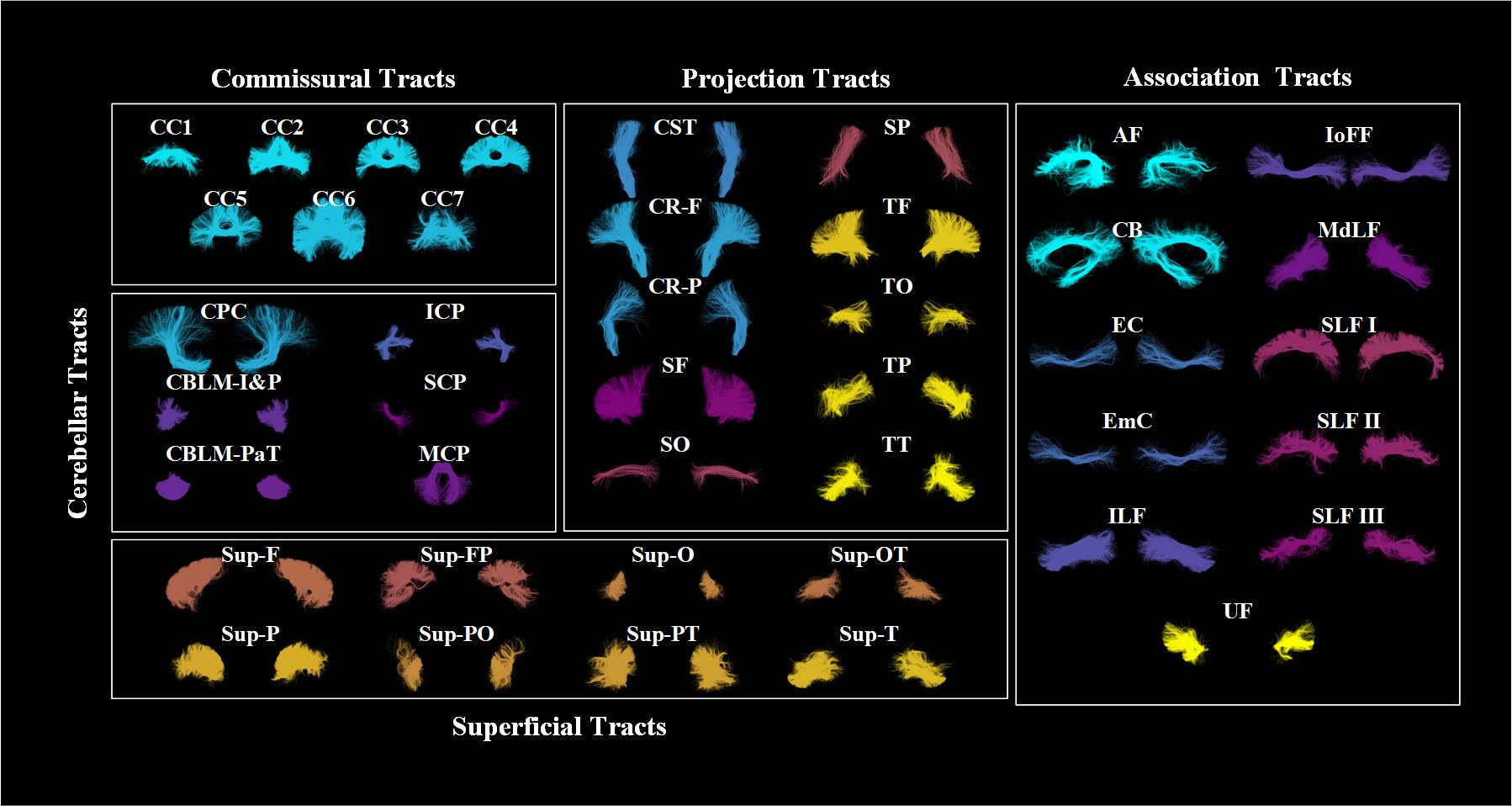}
\caption{Visualization of the anatomical fiber tracts identified in one example subject.}
\label{fig:AtlasOverview}
\end{figure}

With the created WM atlas, we parcellated the whole brain tractography data of each subject in the HCP-YA and CHCP datasets, using the subject-specific fiber clustering tool provided in WMA. In brief, subject-specific tractography data was spectrally embedded into the atlas space by computing the spatial distances of its streamlines to the atlas, followed by the assignment of each fiber to the closest atlas cluster. As a result, the new subject’s tractography was divided into multiple fiber clusters, where each cluster corresponded to a certain atlas fiber cluster. Anatomical tract identification of each subject was conducted by finding the subject-specific clusters that corresponded to the annotated atlas clusters. Figure \ref{fig:AtlasOverview} displays a subject-specific parcellation, demonstrating all anatomical fiber tracts. Then, dMRI measures were extracted from each parcellated cluster and anatomical tract using SlicerDMRI\cite{norton2017slicerdmri,zhang2020slicerdmri} for all subjects. These included the widely used dMRI measures: fractional anisotropy (FA), mean diffusivity (MD), axial diffusivity (AD), radial diffusivity (RD), number of streamlines (NoS), number of streamline points (NoP) and streamline length. 

\section*{Data Records}

Our neuroimaging dataset is now accessible on Science Data Bank\cite{yijie2024adiffusion}, structured following the Brain Imaging Data Structure (BIDS) format\cite{gorgolewski2016brain}. In the root folder, demographic information of the CHCP and HCP-YA cohorts is provided in the “participants.tsv” file, and the harmonized dMRI data and the brain mask of each subject under study are provided in Neuroimaging Informatics Technology Initiative (NIfTI) format. In the 'derivatives' directory, there are two subfolders corresponding to the created WM tractography atlas and the subject-specific WM parcellation results. Specifically, the “derivatives/atlas” folder provides the group-wise fiber clustering atlas including 800 clusters stored in VTK format, the anatomical tract label annotation organized in a 3D Slicer\cite{fedorov20123d} scene file (in medical reality modeling language - MRML, an XML format), and the population mean b0 image by transforming all subjects’ b0 images into the atlas space stored in Neuroimaging Informatics Technology Initiative (NIfTI) format. Instructions about how to visualize the atlas using the 3D Slicer software via SlicerDMRI are provided in a “README.md”. The “derivatives/tractography” provides the whole-brain tractography data, the parcellated fiber clusters, and the anatomical fiber tracts, which are all stored in VTK format. The dMRI measures of each fiber tract (such as FA, MD, and NoS) are also provided in CSV format.  

\section*{Technical Validation}

This section focuses on demonstrating the effectiveness of the dMRI harmonization process, the accuracy of the atlas, and its consistency across diverse populations. Additionally, a preliminary comparison of the CHCP and HCP datasets using the newly proposed atlas is conducted, providing further insights into the applicability of the atlas.

\subsection*{Effects of dMRI Harmonization to Remove Site-Specific Biases}

We first examined the impact of the harmonization procedure. This was done by comparing the voxel-wise diffusion measure FA computed from the dMRI data before and after harmonization at the target site (HCP-YA), in comparison to the reference site (CHCP). The comparison was performed at two different scales, including the entire brain’s WM skeleton and 42 WM regions of interest (ROIs), as delineated by the Illinois Institute of Technology’s (IIT) atlas\cite{zhang2018evaluation,qi2021regionconnect}. All 306 subjects from the CHCP and HCP-YA datasets under study were included in the analysis. The mean FA values between the reference and the original target data and between the original and the harmonized target data were compared. First, independent t-tests were performed to compare the entire brain’s WM skeleton. Next, we performed an ROI-based comparison to assess harmonization’s performance in different regions of the brain. Independent t-tests were performed for each ROI, followed by Bonferroni corrections performed across the multiple comparisons across all ROIs.

\begin{figure}[H]
\centering
\includegraphics[width=0.6\linewidth]{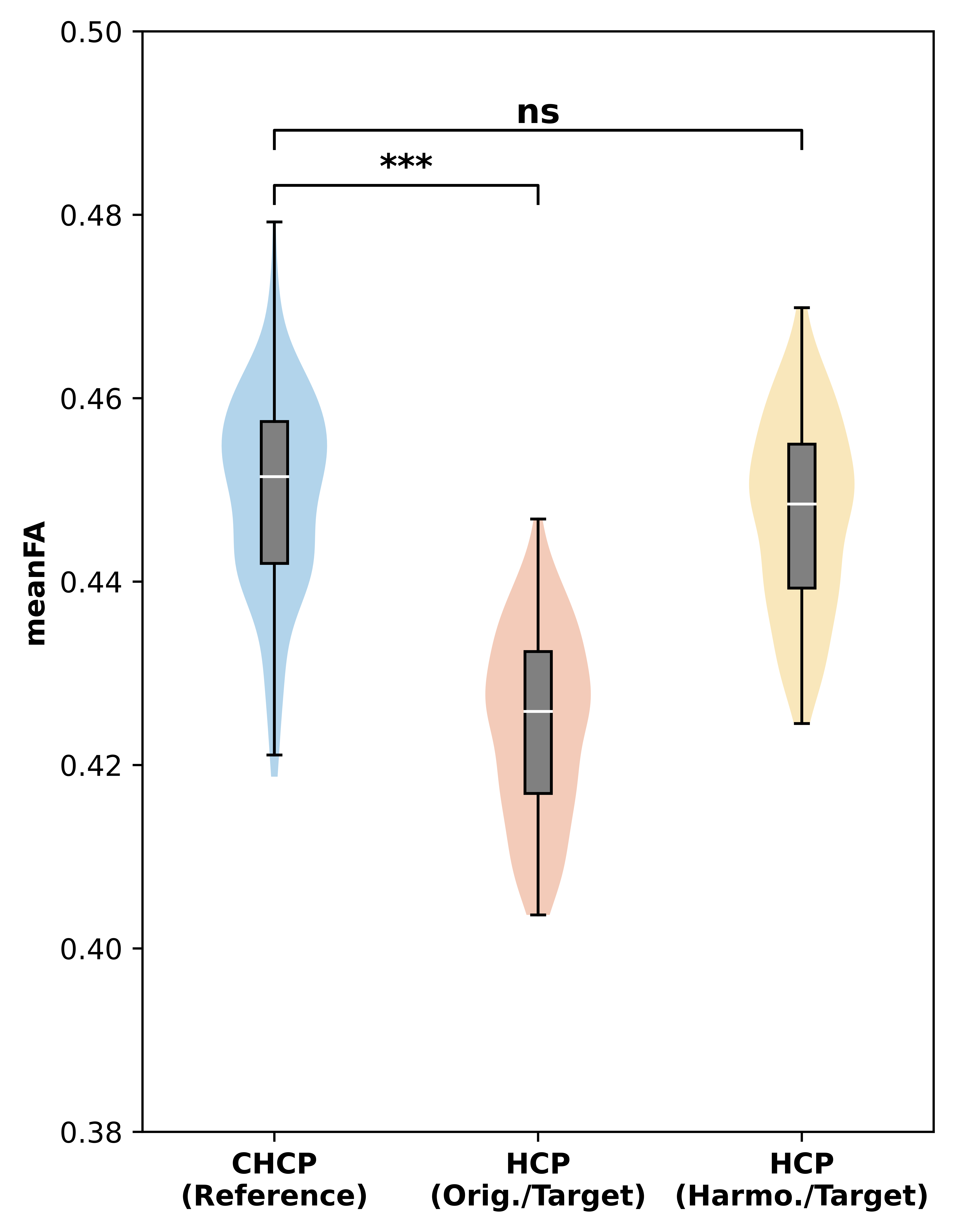}
\caption{Effects of dMRI harmonization in terms of the mean FA of the entire brain’s WM skeleton. Notes: "$ns$" stands for no significance where p-value $>\ 0.05$, $*$ stands for p-value $<\ 0.05$, $**$ stands for p-value $<\ 0.01$, and $***$ stands for p-value $<\ 0.001$.}
\label{fig:SkeletonComparison}
\end{figure}

\begin{figure}[H]
\centering
\includegraphics[width=0.75\linewidth]{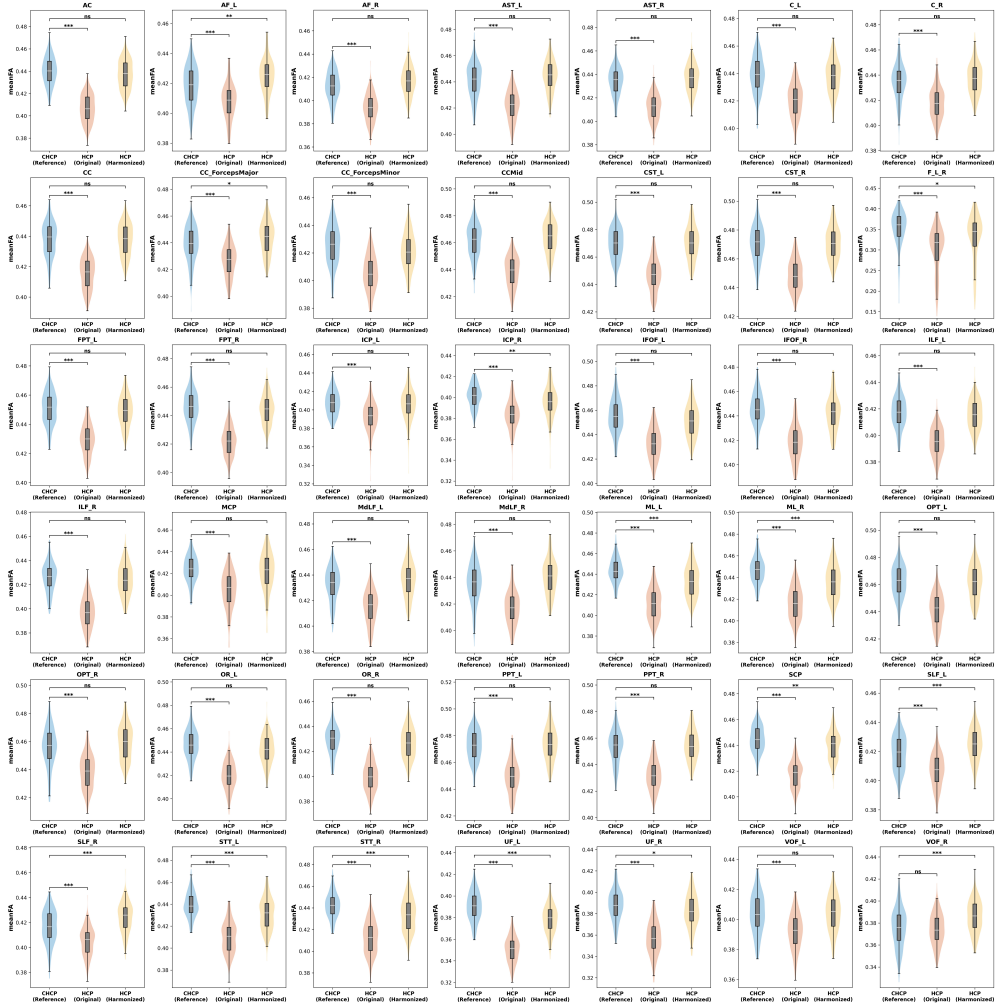}
\caption{Effects of dMRI harmonization on WM ROI analyses. Notes: "$ns$" stands for no significance where p-value $>\ 0.05$, $*$ stands for p-value $<\ 0.05$, $**$ stands for p-value $<\ 0.01$, and $***$ stands for p-value $<\ 0.001$. }
\label{fig:ROIComparison}
\end{figure}

Figure \ref{fig:SkeletonComparison} gives the mean FA over the entire WM skeleton across the CHCP dataset, the original HCP-YA dataset, and the harmonized HCP-YA dataset. We can observe that compared to the original HCP-YA dataset, the FA values of the harmonized HCP-YA dataset became more similar to the CHCP dataset. Notably, the significant differences between the CHCP data and the original HCP-YA data ($p\ll0.01$) were reduced and statistically diminished ($p\ =\ 0.06$) after harmonization. Figure \ref{fig:ROIComparison} gives the comparison results on the 42 ROIs for a more detailed assessment. After harmonization, in all ROIs, the FA values of the harmonized data were closer to those of the reference data, where 14 out of the 42 ROIs showed no significant differences anymore. These indicated that the harmonization process effectively reduced the discrepancies while preserving potential true anatomical variability due to genetic or cultural differences. Overall, the above results underscored the effectiveness of harmonization in eliminating scanner-related differences between the original and reference datasets.

\subsection*{Anatomical Plausibility of Curated Fiber Tracts}
We then assessed the anatomical plausibility of our proposed atlas. This was done by comparing our proposed atlas to the existing ORG atlas\cite{zhang2018anatomically}, which has been previously used in many neuroscientific and clinical research studies\cite{zekelman2022white,cetin2024harmonized,fan2019post,robles2022older,van2024characterization,zanao2023exploring,levitt2023organization}. The evaluation was performed by measuring the spatial similarity of corresponding curated fiber tracts in both atlases using the weighted Dice (wDice) coefficient\cite{zhang2019test,cousineau2017test} — a metric designed for measuring fiber tract spatial overlap. A wDice score over 0.72 is suggested to be a good tract overlap\cite{zhang2019test,cousineau2017test}. 

\begin{figure}[H]
\centering
\includegraphics[width=\linewidth]{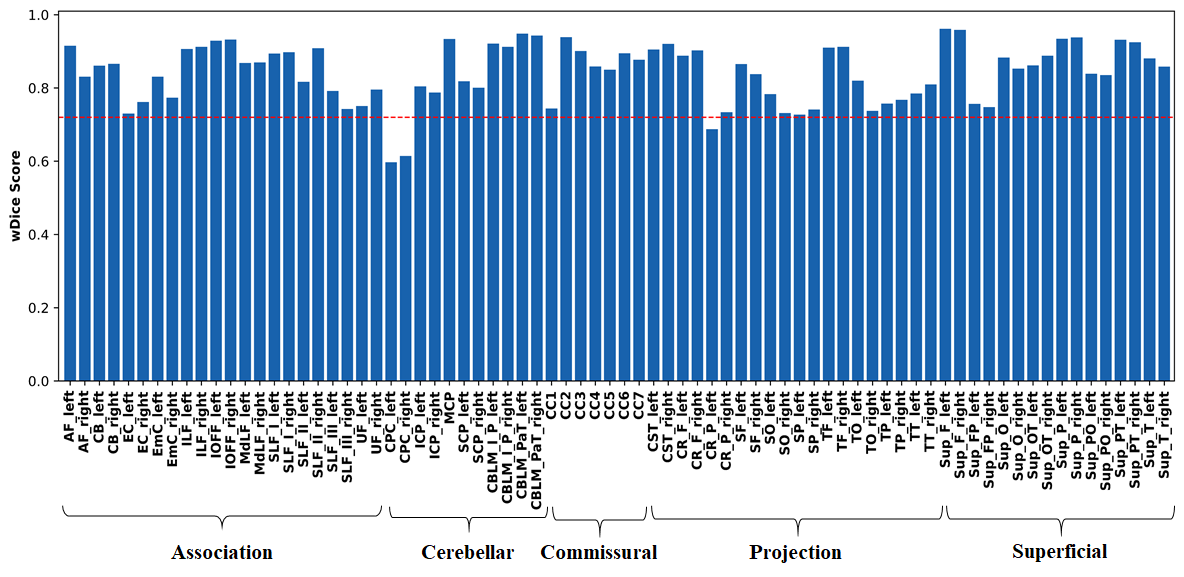}
\caption{wDice score for each of the corresponding anatomical fiber tracts in the ORG atlas and the proposed atlas.}
\label{fig:wDiceScore}
\end{figure}

\begin{figure}[H]
\centering
\includegraphics[width=\linewidth]{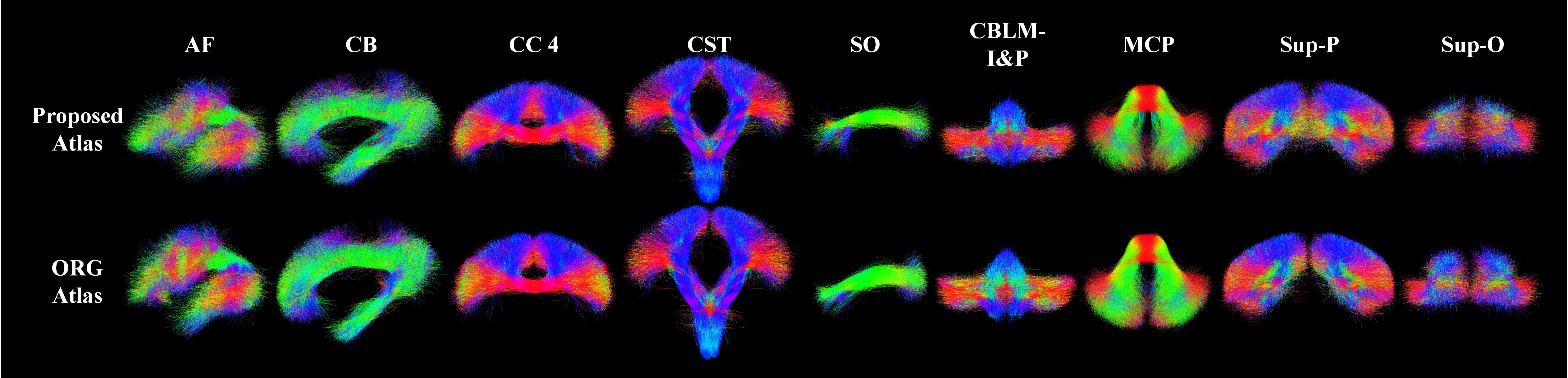}
\caption{Visual comparison of example tracts of the proposed atlas and the ORG atlas.}
\label{fig:AtlasVisualCom}
\end{figure}

Figure \ref{fig:wDiceScore} gives the wDice score for each fiber tract. On average, the mean wDice score across all fiber tracts was 0.84, showing a high volumetric consistency with the ORG atlas. Notably, 73 out of the 76 tracts have a wDice score over 0.72. Figure \ref{fig:AtlasVisualCom} gives a visual comparison of example tracts from the two atlases, where we can observe highly visually comparable tracts between the two atlases. 

\subsection*{Enabling Concurrent WM Mapping Across Populations}

We evaluated the ability of the proposed atlas to enable concurrent mapping of WM tracts in the CHCP and HCP-YA datasets. This was done by comparing it to the ORG atlas that was created purely using the HCP data. To do so, the whole brain tractography of each of the CHCP and HCP-YA subjects was parcellated using the proposed atlas and the ORG atlas separately. Three quantitative metrics were computed to evaluate the WM parcellation performance. The first metric was the WM parcellation generalization (WMPG)\cite{zhang2018anatomically,chen2023deep}, which was measured as the percentage of successfully identified fiber clusters across all subjects in each dataset. A high WMPG value signified strong generalizability of the atlas to a given subject, reflecting the applicability of the atlas across diverse brain structures. The second metric was the inter-subject parcellation variability (ISPV)\cite{zhang2018anatomically,zhang2017comparison}, which was measured as the coefficient of variation (CoV) of NoS for each fiber cluster across each population. A low ISPV value indicated low variability in parcellation, thus signifying a high consistency of WM parcellation across subjects. The third metric was tissue microstructure homogeneity (TMH), which was measured as the CoV of the mean cluster FA values across the subjects in each population\cite{zhang2023ddparcel}. A lower TMH indicates a higher homogeneity of tract microstructural properties as a proxy for better parcellation. 

\begin{figure}[H]
\centering
\includegraphics[width=\linewidth]{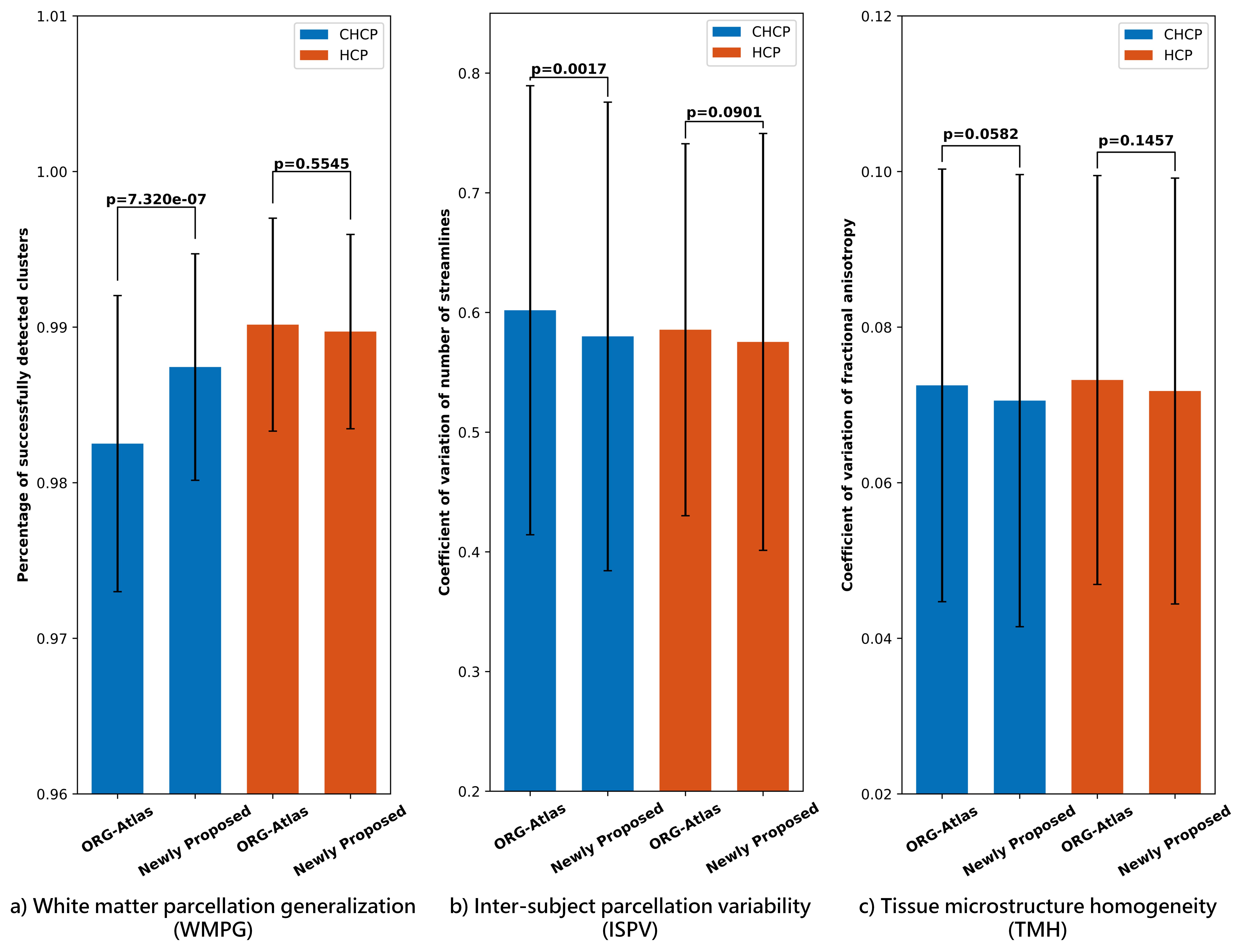}
\caption{Quantitative evaluation of whole brain WM parcellations in the CHCP and HCP-YA populations using the ORG atlas and our newly proposed atlas.}
\label{fig:AtlasQuantityCom}
\end{figure}

Figure \ref{fig:AtlasQuantityCom} presents the quantitative comparison results between employing our proposed atlas and the ORG atlas for subject-specific white matter parcellation. For each metric and each population, an independent t-test was performed to compare the measures between the two atlases. Significant improvement can be observed in the CHCP dataset using the proposed atlas, as indicated by the increased WMPG and decreased ISPV metrics ($p\ \ll\ 0.01$). We can also see a lower TMH using the proposed atlas with a very low p-value ($=\ 0.058$) though without statistical significance. These enhancements demonstrated the effectiveness of incorporating CHCP data alongside the HCP-YA dataset to enhance WM mapping in the CHCP population. On the other hand, it was worth noting that there were no significant differences in any of the three metrics in the HCP-YA data, underscoring the ability of the proposed atlas on concurrent WM mapping across the two populations.

\subsection*{Investigation of WM differences between the Eastern and Western populations}

We first performed a group-wise comparison of the harmonized whole brain tractography data between the CHCP and HCP-YA data. The mean FA value of each subject’s whole brain tractography data was computed, followed by an independent t-test for comparison between the CHCP and HCP-YA datasets. There was no significant difference between the two populations ($p\ =\ 0.70$). (It is worth noticing that there was a significant difference in the whole brain FA before harmonization, $p\ \ll\ 0.01$.) Then, we performed a group-wise comparison for each annotated anatomical fiber tract between the two populations under study. Specifically, for each fiber tract per subject, we extracted two measures of interest, i.e., the percentage of fiber streamlines of the tract in the whole brain tractography to measure the WM connectivity strength and the microstructure measure FA that reflects the diffusion anisotropy of the water molecules of the fiber tract. A general linear model (GLM) analysis was performed for each fiber tract, comparing the CHCP and HCP-YA groups, where the tract measure was the dependent variable, the group was a predictor variable, and age and gender were covariates. A Bonferroni method for multiple comparison corrections was performed across all the fiber tracts. 

\begin{figure}[H]
\centering
\includegraphics[width=0.85\linewidth]{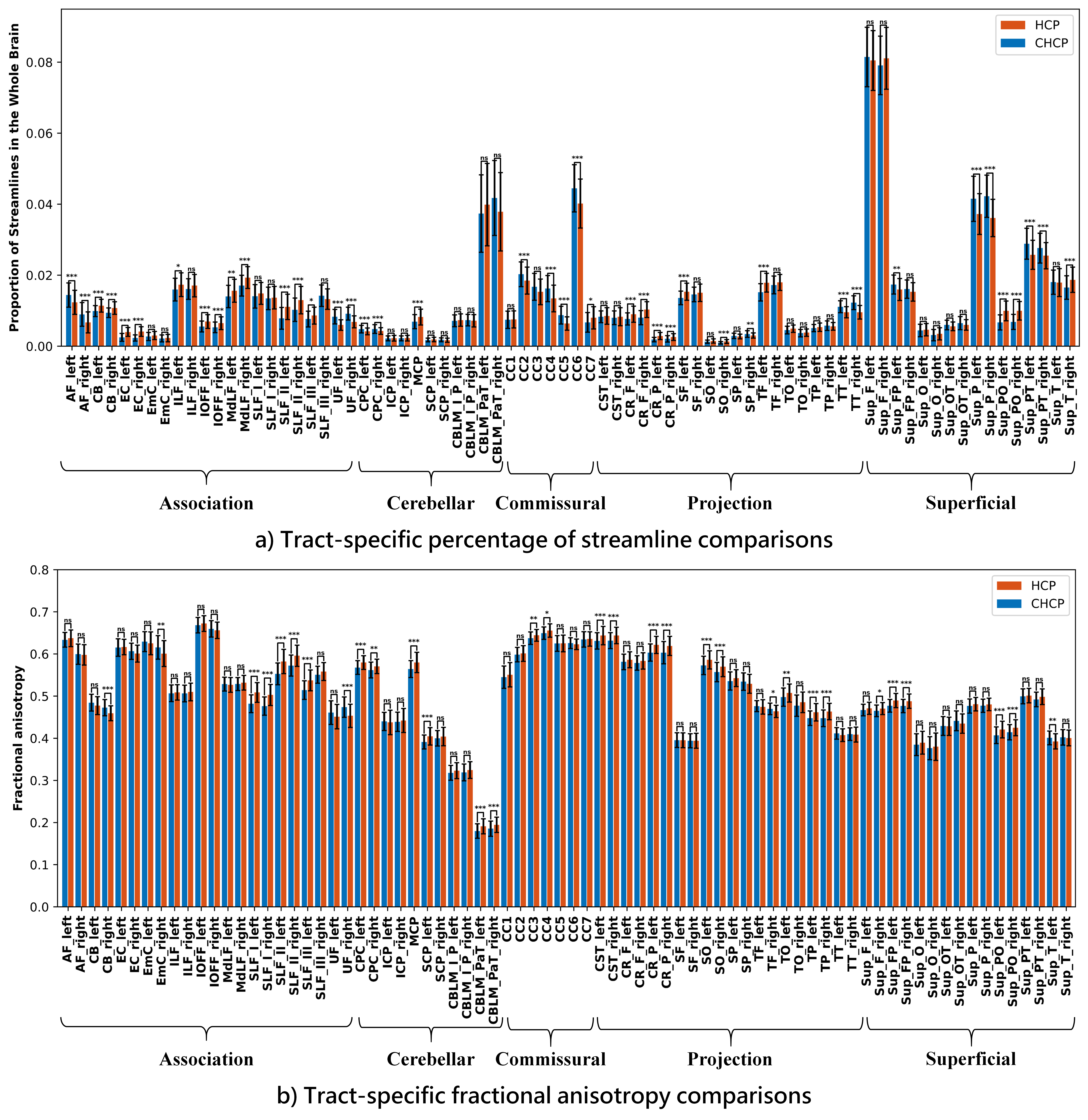}
\caption{(a) shows the comparison of the percentage of streamlines of each anatomical tract between CHCP and HCP-YA datasets. (b) shows the comparison of the FA values for each tract between CHCP and HCP-YA datasets. Notes: "$ns$" stands for no significance where p-value $>\ 0.05$, $*$ stands for p-value $<\ 0.05$, $**$ stands for p-value $<\ 0.01$, and $***$ stands for p-value $<\ 0.001$.}
\label{fig:PopulationComparison}
\end{figure}

Figure \ref{fig:PopulationComparison} shows the results of the comparison between the two populations. In the connectivity strength comparison, there were 42 tracts with significant differences, in which 18 tracts had higher strengths in the CHCP group and 24 tracts had higher strengths in the HCP-YA group. In the FA comparison, there were 32 tracts with significant differences, in which 5 tracts had higher FA values in the CHCP group and 27 tracts had higher FA values in the HCP-YA group. 

\section*{Usage Notes}

In this study, we created a dMRI tractography atlas to enable concurrent mapping of WM connections in both Eastern and Western populations. Alongside the atlas, we also generated a large harmonized dMRI dataset from the two populations, including the diffusion imaging data, the whole brain tractography data, and parcellated WM fiber tracts and their diffusion measures. This resource serves as a valuable tool for exploring both commonalities and differences among diverse cultural backgrounds, offering openly accessible data to support further investigations. The potential usage of our work is discussed in the following aspects.

The subject-specific fiber tract data can be a valuable resource for conducting cross-cultural investigations into local WM regions associated with certain brain functions. These fiber tract data were derived from the harmonized dMRI dataset and the proposed cross-population atlas as a common template, enabling an unbiased comparison of corresponding WM structures between the two populations. Our initial exploration of the data shows significant differences between the two populations in several language-related fiber tracts such as AF, IOFF, UF, MdLF, SLF II, and ILF. This finding is consistent with the observed differences in language processing tasks between these cultural groups, as identified through functional MRI analyses\cite{ge2023increasing,kochunov2003localized,ge2015cross,wei2023native,lu2021functional,wu2015direct}. Further investigation into fiber tracts related to other brain functions, such as motor and cognition, can elucidate potential differences in brain function, providing insights into the variations that may exist across cultural groups. 

In addition to tract-specific analyses, the whole brain tractography data is available for conducting connectome-style analyses. The harmonized tractography data can serve as a valuable tool for further validating the results obtained in previous multi-site studies\cite{ge2023increasing,zhang2019structural,suo2021anatomical}, eliminating the need for additional data harmonization processes. By providing a standardized and consistent dataset across populations, the harmonized tractography enables a more direct comparison and evaluation of findings from diverse study sites. This contributes to the robustness and reliability of existing results, enhancing the overall validity of cross-cultural comparisons in brain research.

Moreover, the harmonized dMRI dataset empowers researchers to conduct a wide range of analyses to test their neuroscientific hypotheses concerning cross-cultural brain differences in dMRI. Unlike existing dMRI harmonization approaches that utilize meta-analysis or mega-analysis to aggregate data from different sites\cite{kochunov2014multi,zhang2016enigma}, our method harmonizes the dMRI data at the diffusion signal level. Beyond the aforementioned tractography-based analysis, the provided dMRI data can be seamlessly integrated into various downstream analysis methods. These include voxel-wise tract-based spatial statistics (TBSS)\cite{smith2006tract}, voxel-based morphometry (VBM)\cite{ashburner2000voxel} on standard diffusion tensor features, and advanced models like Neurite Orientation Dispersion and Density Imaging (NODDI)\cite{zhang2012noddi} and Free Water (FW)\cite{pasternak2009free}. 

Lastly, the generated atlas and associated computational tools are publicly accessible, facilitating automated parcellation of WM tracts in new datasets. Through the harmonization of the new data to the atlas dMRI dataset, researchers can conduct WM parcellation based on the atlas. In this context, the proposed atlas serves as a valuable resource for cross-cultural comparisons in the context of brain diseases, aging, and neurodevelopment. The availability of this tool enhances the potential for comprehensive investigations into structural brain variations across diverse cultural contexts.

\section*{Code availability}

The dMRI data processing software employed in this study is publicly available on GitHub. The repositories house the complete processing pipeline and scripts covering stages from preprocessing, CNN-based masking, dMRI harmonization, to whole-brain tractography, white matter parcellation, and visualization. These repositories offer comprehensive documentation, usage instructions, script examples on sample data, and list all dependencies necessary for effective utilization of the software.

\begin{itemize}
\item CHCP dMRI data processing pipeline: \url{https://github.com/pnlbwh/pnlpipe/tree/v2.2.0}
\item Convolutional neural network-based dMRI brain segmentation: \\ \url{https://github.com/pnlbwh/CNN-Diffusion-MRIBrain-Segmentation/tree/v0.3}
\item dMRI data harmonization: \url{https://github.com/pnlbwh/dMRIharmonization/tree/v2.1}
\item UKF whole brain tractography: \url{https://github.com/pnlbwh/ukftractography}
\item White Matter Analysis: \url{https://github.com/SlicerDMRI/whitematteranalysis/tree/WMA-Faster}
\item SlicerDMRI: \url{http://dmri.slicer.org}
\end{itemize}

\section*{Acknowledgments}

This work is in part supported by the National Key R\&D Program of China (No. 2023YFE0118600), the National Natural Science Foundation of China (Nos. 62371107, 62201265), and the National Institutes of Health (R01MH125860, R01MH119222, R01MH132610, R01NS125781).  

\bibliography{main}

\end{document}